\documentclass[journal]{IEEEtranTIE}
\usepackage{graphicx}
\graphicspath{{figures/}}
\usepackage{cite}
\usepackage{picinpar}
\usepackage{amsmath}
\usepackage{mathptmx}
\usepackage{mathtools}
\usepackage{url}
\usepackage{flushend}
\usepackage[latin1]{inputenc}
\usepackage{colortbl}
\usepackage{soul}
\usepackage{multirow}
\usepackage{pifont}
\usepackage{color}
\usepackage{alltt}
\usepackage[hidelinks]{hyperref}
\usepackage{enumerate}
\usepackage{siunitx}
\usepackage{breakurl}
\usepackage{epstopdf}
\usepackage{pbox}
\usepackage{cancel}
\usepackage[table]{xcolor}
\usepackage{tcolorbox}
\definecolor{chired}{RGB}{173, 62, 46}
\definecolor{reddish}{RGB}{255, 230, 183}
\allowdisplaybreaks

\begin{document}
\title{LCL Resonance Analysis and Damping in Single-Loop Grid-Forming Wind Turbines}

\author{
	\vskip 1em
	
	Meng Chen, \emph{Member, IEEE},
	Yufei Xi, \emph{Member, IEEE},
    Frede Blaabjerg, \emph{Fellow, IEEE},
    Lin Cheng, \emph{Senior Member, IEEE},
	and Ioannis Lestas, \emph{Member, IEEE}

	\thanks{
	
		This work was funded by UK Research and Innovation (UKRI) under the UKRI Postdoctoral Fellowship Guarantee Grant [EP/Z001889/1]. M. Chen would like to acknowledge the Department of Engineering, University of Cambridge, for the support of the Fellowship. The contents reflect only the authors' view and not the views of the University or the UKRI. (\textit{Corresponding author: Yufei Xi.})

Meng Chen and Ioannis Lestas are with the Department of Engineering, University of Cambridge, Cambridge CB2 1PZ, United Kingdom (e-mail: mc2545@cam.ac.uk; icl20@cam.ac.uk).

Yufei Xi and Lin Cheng are with the State Key Laboratory of Power System Operation and Control (Department of Electrical Engineering, Tsinghua University), Beijing 100084, China (e-mail: xyf.2022@tsinghua.org.cn; chenglin@mail.tsinghua.edu.cn).

Frede Blaabjerg is with the AAU Energy, Aalborg University, 9220 Aalborg, Denmark (e-mail: fbl@energy.aau.dk).
	}
}

\maketitle
	
\begin{abstract}
A common assumption in both grid-following (GFL) and grid-forming (GFM) control systems is that they are open-loop (OL) stable in the vicinity of high-frequency resonances. Hence classical loop-shaping approaches are often used for establishing stability margins and designing active damping (AD) strategies. This paper shows that single-loop GFM (SL-GFM) control schemes incorporating a widely used class of reactive power (RAP) control, referred to as droop-I control, can lead to OL unstable poles. This finding reveals a novel instability mechanism resulting in a reduced stability margin and robustness at high frequencies. The sensitivity of this phenomenon to both RAP and electrical parameters is analyzed in detail. An AD design that explicitly accounts for the newly identified instability mechanism is proposed. We also provide a comparison  between such SL-GFM and well-studied GFL control schemes, highlighting quite different resonance features between them. Validation is performed through experiments.
\end{abstract}



\definecolor{limegreen}{rgb}{0.2, 0.8, 0.2}
\definecolor{forestgreen}{rgb}{0.13, 0.55, 0.13}
\definecolor{greenhtml}{rgb}{0.0, 0.5, 0.0}

\section{Introduction}

\IEEEPARstart{R}{enewable} energy generators are increasingly equipped with grid-forming (GFM) control strategies to enable grid support functions \cite{Chen2024}. From a control loop perspective, GFM control can be broadly categorized into single-loop (SL), double-loop, and multi-loop structures \cite{Pan2020,Du2020,Chen2024,Liu2024a}. The SL structure does not include explicit current control for limiting fault currents. Nevertheless, alternative approaches such as adaptive virtual impedance methods \cite{Wu2023} and power reference regulation methods \cite{Liu2022} have been proposed for limiting fault currents while remaining the operation in GFM mode. Meanwhile, compared to other control structures, the SL structure is simple for modeling and analysis. Moreover, due to the absence of  fast inner control loops, it is less prone to instabilities caused by high sensitive to time delays \cite{Akhavan2023} and by insufficient bandwidth separation among multiple control loops \cite{Li2025}. In addition, it has been shown that the SL structure exhibits improved small-signal stability properties as the filter effectively softens the dynamic coupling between the GFM converters and the grid \cite{Du2020}. Despite these differences, droop-based control usually remains a common feature.

In the context of reactive power (RAP) control, two primary droop-based strategies are commonly used. The first is conventional droop control, where the filtered RAP is fed back to generate the voltage reference directly \cite{Liu2022}. However, it cannot provide robust droop characteristics in the SL structure and is influenced by the filter parameters, operating currents, and additional blocks such as virtual impedance. This issue can be effectively addressed using the second approach, referred to in this paper as droop-I control. The droop-I control feeds back both RAP and voltage magnitude, using an integrator to generate the voltage reference \cite{Pan2020}. Owing to the integrator action, the droop-I control has been widely used to provide robust droop characteristics \cite{Zhao2022,Wu2023,Sun2023}.

Meanwhile, the stability analysis of GFM control can be conducted using either simplified or more detailed models. Simplified models typically assume stable fast dynamics, focusing instead on the low-frequency power loops \cite{Ojo2020,Liu2025}, where bandwidth is generally lower than the synchronous frequency of 50 Hz \cite{Du2020}. Such analysis has remained a central focus in GFM control research and has led to substantial progress in power loop stability\cite{Rosso2021, Chen2022}. In contrast, detailed models explicitly capture faster dynamics such as synchronous resonances \cite{Xiong2023} and high-frequency resonances introduced by LCL filters. Depending on the filter parameters, the high-frequency resonant frequencies typically range from several hundred Hertz to above 1 kHz\cite{Liu2024a}.

LCL resonance has been extensively investigated in grid-following (GFL) control, particularly with respect to its interaction with the current control loop and the design of active damping (AD) strategies \cite{Parker2014,Wang2016,Liu2020,Wang2022}. More recently, similar analyses have been extended to SL-GFM converters \cite{Liu2024a}. A conventional assumption underlying these studies is that the system exhibits open-loop (OL) stable characteristics near high-frequency resonance when time delays are neglected. Based on this condition, classical loop shaping approaches and gain-phase crossover conditions \cite{Franklin2019} remain a widely adopted method for assessing stability margins and guiding AD design.

While prior studies have used detailed models that incorporate high-frequency dynamics \cite{Du2020,Liu2024a}, they have primarily focused on conventional droop control. It is not clear whether the SL-GFM converters with droop-I control have the same characteristics with respect to the high-frequency resonance. Meanwhile, if new resonant characteristics are found, the corresponding stability assessment framework and AD design may need to be revised.

This paper shows that SL-GFM control schemes incorporating droop-I control may become nonminimum phase due to the presence of OL unstable poles (As the number of OL zeros have nothing to do with the closed-loop (CL) stability, the terms minimum phase and nonmininum phase in this paper only consider the poles.), where such nonminimum phase behavior has not been observed in previous research. This is a significant departure from conventional modeling assumptions that leads to very different high-frequency resonance characteristics, making many previously established control designs inapplicable with poor stability margins.

The key contributions of this paper are summarized as follows:
\begin{enumerate}
    \item A new instability mechanism is identified in which the RAP OL system transitions from minimum phase to nonminimum phase due to the emergence of OL unstable poles. 
    \item The coupling between droop-I control and high-frequency resonance modes is revealed, highlighting the different resonance behavior of the droop-I SL-GFM converters.
    \item Different RAP control strategies are compared to evaluate their respective effects on high-frequency resonant modes.
    \item A step-by-step AD design method based on capacitor voltage feedback is proposed for implementation in SL-GFM converters. 
    \item A comprehensive comparison between droop-I SL-GFM and well-studied GFL control is presented, highlighting their fundamentally different resonance characteristics and the resulting AD design. 
\end{enumerate}

The remainder of the paper is organized as follows: Section \ref{sec_HF} presents a comprehensive small-signal analysis of the SL-GFM converter, emphasizing the significance of LCL resonance. Section \ref{sec_AD} details the AD design methodology in SL-GFM converters. Section \ref{sec_comparison} summarizes a comparison with GFL converters. Section \ref{sec_test} provides experimental results, while Section \ref{sec_con} summarizes the key findings and conclusions.

\section{High-Frequency Resonance in Single-Loop Grid-Forming Converters}\label{sec_HF}

\subsection{Structure and Conventional Simplified Model}

\begin{figure*}[!t]
\centering
\includegraphics[width=\textwidth]{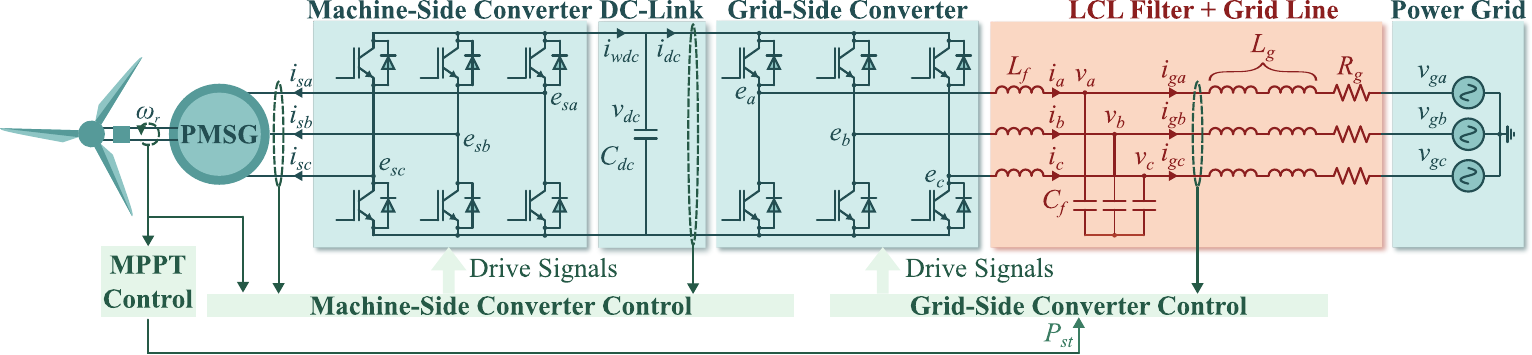}
\caption{Structure of a PMSG-WT system with LCL filter controlled by SL-GFM control strategy.}
\label{fig_structure}
\end{figure*}

\begin{figure}[!t]
\centering
\includegraphics[width=\columnwidth]{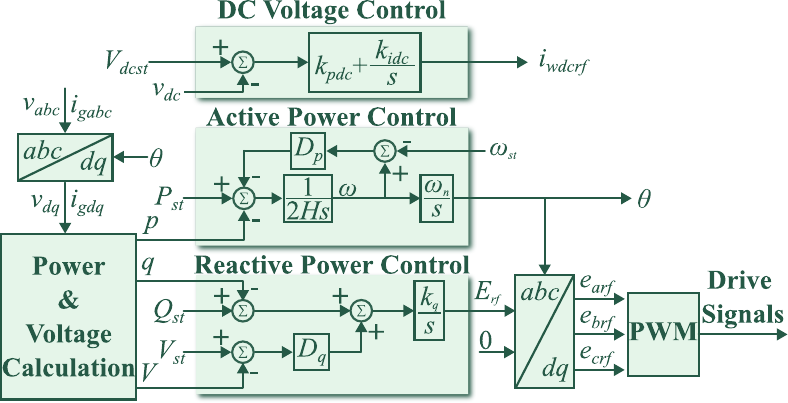}
\caption{Detailed control block diagram of PMSG-WT system with droop-I RAP control.}
\label{fig_control}
\end{figure}

We assume the SL-GFM control is incorporated in a permanent magnet synchronous generator-based wind turbine (PMSG-WT) system shown in Fig. \ref{fig_structure}. The control block diagrams shown in Fig. \ref{fig_control} and the analysis are implemented in p.u. values.

The DC-link voltage is regulated using a proportional-integral (PI) controller, which is expressed as \cite{Chen2022}:
\begin{align}
    \label{eq_iwdcrf}
    &i_{wdcrf} = k_{pdc}(V_{dcst} - v_{dc}) + k_{idc}\int(V_{dcst} - v_{dc})dt
\end{align}
where $i_{wdcrf}$ is the reference DC current from the machine-side converter (MSC), $k_{pdc}$ and $k_{idc}$ are the proportional and integral gains, respectively, $v_{dc}$ is the DC-link voltage, and the subscript "st" represents the set-point value. By simplifying the MSC as a controlled current source, it is assumed that the DC current from the MSC $i_{wdc}$ equals its reference, i.e., $i_{wdcref}=i_{wdc}$ \cite{Chen2022}. Meanwhile, assuming the grid-side converter (GSC) is lossless, the input and output powers are equal. The DC-link voltage dynamics with AC and DC sides coupling is:
\begin{align}
    \dot v_{dc} = (\omega_n/C_{dc})i_{wdc} - (\omega_n/C_{dc})E_{rf}i_d/v_{dc}
\end{align}
where $\omega_n$ is the nominal angular frequency, $C_{dc}$ is the DC-link capacitance, $i$ is the current of filter inductor $L_f$, $E_{rf}$ is the voltage magnitude of the SL-GFM converter, and the subscript "$d$" and "$q$" indicate the $d$-axis and $q$-axis components of corresponding variables.

\begin{figure}[!t]
\centering
\includegraphics[width=\columnwidth]{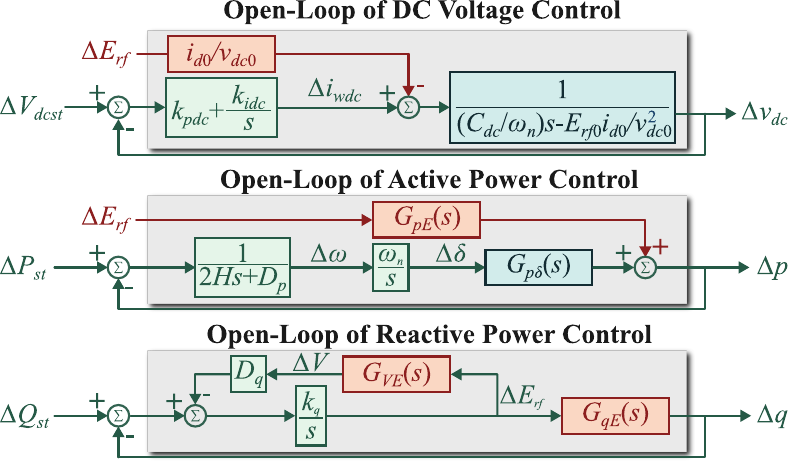}
\caption{Small-signal block diagram of SL-GFM PMSG-WT system.}
\label{fig_small_signal}
\end{figure}

On the AC side, the SL-GFM derives voltage references with magnitude $E_{rf}$ and angle $\theta$ \cite{Oraa2024}. It is noticed that in addition to the swing equation-based active power (AP) control, the RAP control uses the droop-I control \cite{Sadeque2024}: 
\begin{align}
\label{eq_DroopI}
    (1/k_q)\dot E_{rf} = Q_{st} - q + D_q(V_{st} - V)
\end{align}
where, due to the integrator, the right-hand side of (\ref{eq_DroopI}) becomes zero in steady-state, i.e., $Q_{st} - q + D_q(V_{st} - V) = 0$. Therefore, the $q$-$V$ droop characteristic is achieved, with the droop coefficient given by $D_q$. This relationship is independent of system parameters and is not affected by additional blocks such as virtual impedance \cite{Abdelghany2025,Xia2025}. The integrator gain $k_q$ determines the response speed of the system as it approaches the droop steady-state. Thus, the small-signal block diagram of the system can be derived as shown in Fig. \ref{fig_small_signal}.

In conventional simplified models, the high-frequency dynamics of the LCL filter are neglected. Therefore, the corresponding transfer functions in Fig. \ref{fig_small_signal} can be given by \cite{Chen2021}:
\begin{align}
    &G_{p\delta}(s) \coloneqq (E_{rf0}V_g/X_{eq})\cos\delta_0\\
    \label{eq_Gqe_sim}
    &G_{qE}(s) \coloneqq [2E_{rf0}X_g + V_g\cos\delta_0(2X_f - X_{eq})]/X^2_{eq}\\
    \label{eq_Gve_sim}
    & G_{VE}(s) \coloneqq \frac{X_g(E_{rf0}X_g + V_gX_f\cos\delta_0)}{X_{eq}\sqrt{E^2_{rf0}X^2_g + 2E_{rf0}V_gX_fX_g\cos\delta_0 + V^2_gX^2_f}}
\end{align}
where the subscript "0" denotes the equilibrium point, $X$ represents the static impedance, and $\delta$ is the angle difference between voltages of GSC and grid, expressed by
\begin{equation}
    \label{eq_delta}
    \dot\delta = \omega_n(\omega - \omega_g)
\end{equation}
where $\omega_g$ is the grid angular frequency. Meanwhile, we have $X_{eq} \coloneqq X_f + X_g - X_cX_fX_g$. It is noticed in Fig. \ref{fig_structure} that the LCL grid-side inductance and the grid equivalent inductance are combined into a single inductance $L_g$ for convenience \cite{Liu2025}. Neglecting the loop coupling leads to simplified OL transfer functions (TFs) for the DC loop, AP loop, and RAP loop:
\begin{align}
    \label{eq_Gdc_sim}
    &G_{dc\_sim}(s) \coloneqq \left(k_{pdc} + \frac{k_{idc}}{s}\right)\frac{\omega_n}{C_{dc}s + \omega_nG_{dcv}(s)}\\
    &G_{p\_sim}(s) \coloneqq \frac{1}{2Hs + D_p}\frac{\omega_n}{s}G_{p\delta}(s)\\ 
    \label{eq_Gq_sim}
    &G_{q\_sim}(s) \coloneqq G_{qE}(s)/[(1/k_q)s + D_qG_{VE}(s)]
\end{align}

\begin{figure}[!t]
\centering
\includegraphics[width=\columnwidth]{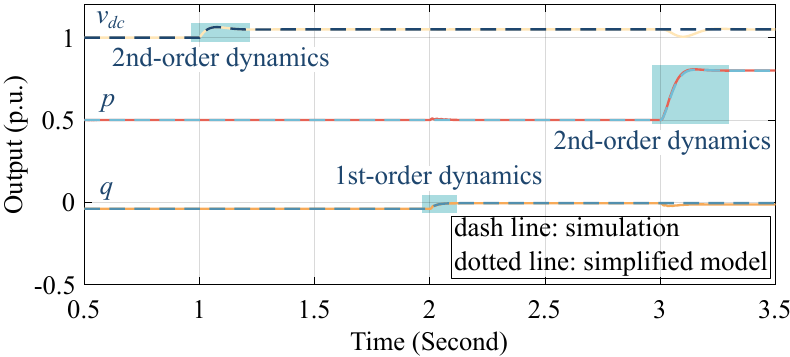}
\caption{Comparison of step responses between simplified small-signal model and detailed model of time-domain simulation.}
\label{fig_step}
\end{figure}

The simplified and decoupled OL models (\ref{eq_Gdc_sim})-(\ref{eq_Gq_sim}) are directly derived from Fig. \ref{fig_small_signal} and are typically used in GFM control analysis. According to Nyquist Criterion, these OL models can be used to evaluate the corresponding CL stabilities. With the parameters in Table \ref{tab_parameter}, their step responses are compared with a detailed model simulation, where $V_{dcst}$, $P_{st}$, and $Q_{st}$ step 0.05 p.u. at $t=1$ s, 0.3 p.u. at $t=2$ s, and 0.1 p.u. at $t=3$ s, from 1 p.u., 0.5 p.u., and 0 p.u., as shown in Fig. \ref{fig_step}. In Table \ref{tab_parameter}, the droop coefficients $D_p$ and $D_q$ are selected according to typical droop regulation requirements, i.e., a 2\% frequency deviation results in a 100\% change in output AP and a 10\% voltage deviation leads to a 100\% change in output RAP \cite{Wu2023}. $H$, $k_{pdc}$, and $k_{idc}$ are selected to ensure favorable dynamic performance with damping ratios of 0.707 for both the AP and DC-link voltage modes. $C_f$ is selected so that the power factor variation less than 5\%. The ratio between the inverter-side and grid-side filter inductors is set to 1 with the filter resonant frequency to be 1 kHz \cite{Jayalath2017}. Conventionally, it is assumed that the low-frequency power modes are dominant and the high-frequency modes can be neglected. Therefore, the simplified model is sufficient. Fig. \ref{fig_step} also shows that the responses of the simplified models seem to closely match the simulation dynamics. However, in the next section we will show that such assumption may be not proper and the simplified model may not reflect the dominant stability feature of the overall system by revealing that:
\begin{enumerate}
    \item The system exhibits a small stability margin dominated by LCL resonance modes and is prone to instability, a behavior not captured by the simplified model.
    \item RAP control is intrinsically coupled with the LCL resonance modes, limiting the ability to freely tune the RAP dynamics as assumed in the simplified model.
\end{enumerate}

\begin{table}[!t]
    \renewcommand{\arraystretch}{1.3}
    \centering
    \caption{Parameters of SL-GFM PMSG-WT System}
    \resizebox{\columnwidth}{!}{
    \begin{tabular}{clc}
        \hline\hline\\[-3mm]
        Parameters & \multicolumn{1}{c}{Description} & Values \\ \hline
        $V_{dcn}$ & Nominal DC voltage & 1200 V (1 p.u.) \\ 
        $f_{sw}$ & Switching frequency & 5 kHz \\
        $C_f$ & Filter capacitance & 1.6 mF (0.048 p.u.)\\
        $L_f$ & Inverter-side filter inductance & 32 $\mu$H (0.1 p.u.)\\
        $L_g$ & Grid-side filter + line inductance & 60 $\mu$H (0.2 p.u.)\\
        $X_g/R_g$ & $X/R$ ratio of the line & 6\\
        $C_{dc}$ & DC-link capacitance & 0.3 F (27 p.u.)\\
        $H$ & Inertia constant & 0.5 s\\
        $D_p$ & Damping of active power control & 50\\
        $D_q$ & Droop coefficient of $q-V$ control & 10\\
        $k_q$ & Integral gain of $q-V$ control & 4\\
        $\omega_{st}$ & Set-point of angular frequency & 1 p.u.\\
        $Q_{st},~V_{st}$ & Set-points of reactive power control & 0 p.u., 1 p.u.\\
        $V_{dcst}$ & Set-point of DC-link voltage & 1 p.u.\\
        $k_{pdc},~k_{idc}$ & Gains of DC voltage PI & 3.8, 63.8\\[1.4ex]
        \hline\hline
    \end{tabular}
    }
    \label{tab_parameter}
\end{table}

\subsection{Analysis on High-Frequency Dynamics}\label{subsec_DetailedModel}

To investigate the effects of the LCL high-frequency dynamics, detailed models should be used as given in \cite{Chen2022}:
\begin{align}
    \label{eq_ded}
    & \tau_d\dot e_d = E_{rf} - e_d\\
    \label{eq_did}
    & (L_f/\omega_n)\dot i_d = e_d - v_d + \omega L_fi_q\\
    & (L_f/\omega_n)\dot i_q = -v_q - \omega L_fi_d\\
    & (C_f/\omega_n)\dot v_d = i_d - i_{gd} + \omega C_fv_q\\
    & (C_f/\omega_n)\dot v_q = i_q - i_{gq} - \omega C_fv_d\\
    & (L_g/\omega_n)\dot i_{gd} = v_d - V_gcos\delta - R_gi_{gd} + \omega L_gi_{gq}\\
    \label{eq_digq}
    & (L_g/\omega_n)\dot i_{gq} = v_q + V_gsin\delta - R_gi_{gq} - \omega L_gi_{gd}.
\end{align}
where (\ref{eq_ded}) estimates the delay $e^{-\tau_ds}$ introduced by switching behavior and digital control and is approximated as 1.5 times the sampling period $T_s$. The model only considers the grid-side resistance while the parasitic resistances of other branches are neglected. This is because these parasitic resistances are typically much smaller than the grid resistance. Meanwhile, it is well established that the resistances in the LCL filter branches, as dissipative elements, inherently provide damping to the resonance. Thus, neglecting them corresponding to a worst-case scenario, where the inherent damping is minimized. 

\begin{figure}[!t]
\centering
\includegraphics[width=\columnwidth]{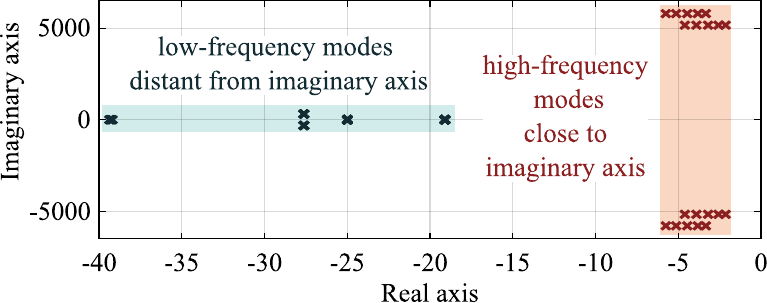}
\caption{Eigenvalues of detailed model of SL-GFM PMSG-WT system with $\tau_d\in[0, 1.5T_s]$.}
\label{fig_pole}
\end{figure}

By combining (\ref{eq_iwdcrf})-(\ref{eq_DroopI}), (\ref{eq_delta}), and (\ref{eq_ded})-(\ref{eq_digq}), the detailed system model is formulated. Considering a sampling frequency of 10 kHz, its eigenvalues are shown in Fig. \ref{fig_pole} for different values of $\tau_d\in[0, 1.5T_s]$. The step responses in Fig. \ref{fig_step} are primarily influenced by low-frequency modes, which are located far from the imaginary axis, indicating a large stability margin. It is noted that Fig. \ref{fig_pole} also presents two resonance modes, whose dynamics are not captured in the simplified small-signal model nor observed in the time-domain simulation due to their high frequency. However, these high-frequency modes are located close to the imaginary axis, suggesting a small stability margin and potential risk of instability under practical parameter variations. Fig. \ref{fig_pole} also indicates that $\tau_d$ has a limited impact on high-frequency stability since the SL-GFM control does not include a fast current loop. Therefore, in the following analysis, $\tau_d$ is assumed to be compensated and is neglected. A detailed discussion of the compensation method can be found in \cite{Liu2020}. Meanwhile, since sensor noise does not change the intrinsic characteristics of the system, it is assumed that the measuring is ideal \cite{Liu2024a}.

\begin{figure}[!t]
\centering
\includegraphics[width=\columnwidth]{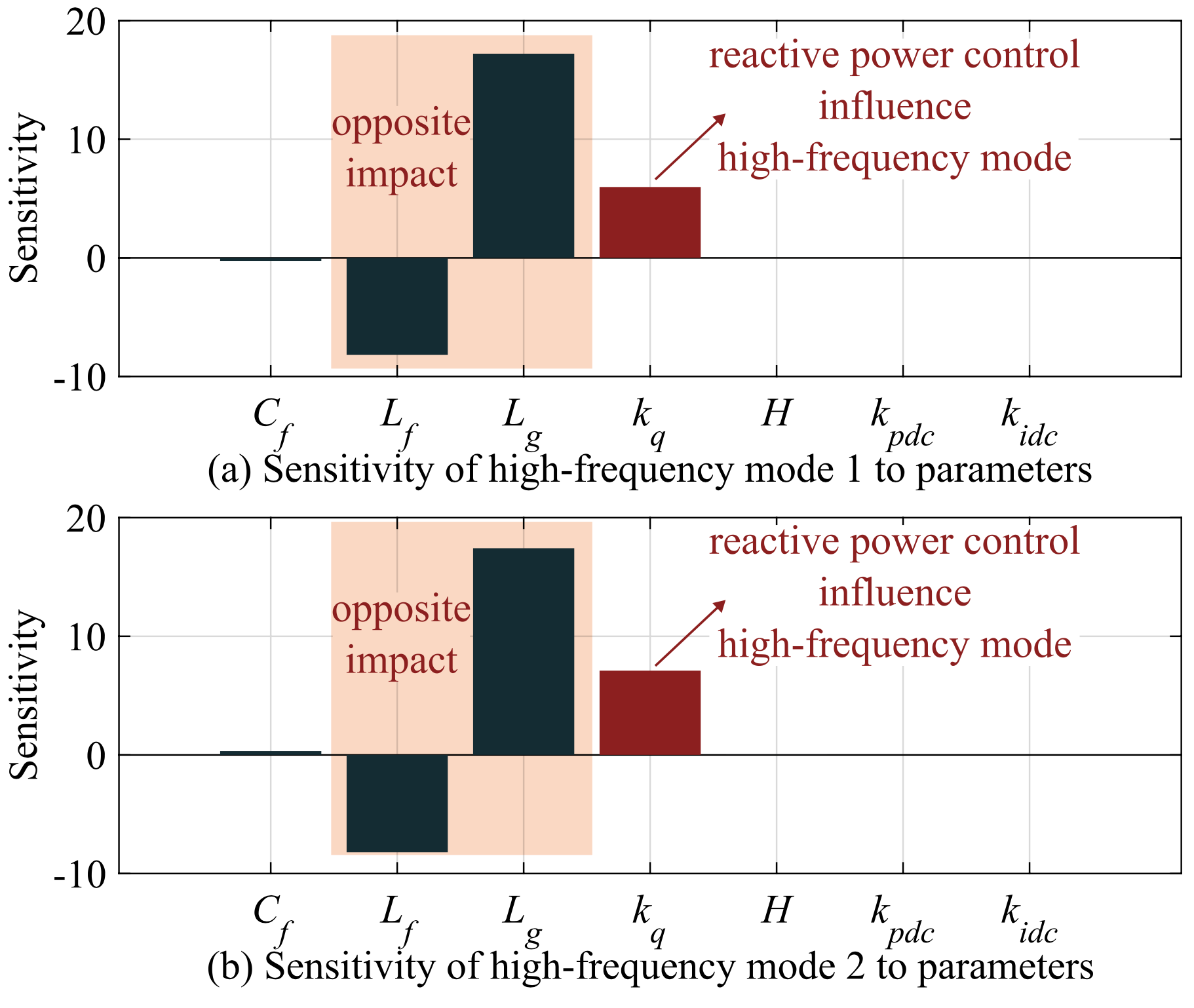}
\caption{Sensitivity evaluating variation of real parts of resonance modes in response to percentage changes in system parameters.}
\label{fig_sensitivity}
\end{figure}

A sensitivity analysis \cite{Franklin2019} is conducted to evaluate the variation of the real parts of the resonance modes in response to percentage changes in system parameters. The control parameters associated with the AP and DC-link voltage control loops mainly influence the low-frequency dynamics. The results, presented in Fig. \ref{fig_sensitivity}, verify that they have negligible influence on the high-frequency resonant modes. More importantly, Fig. \ref{fig_sensitivity} reveals two key observations:
\begin{enumerate}
    \item Apart from the LCL filter parameters, RAP control significantly affects the high-frequency modes, challenging conventional assumptions.
    \item The inverter-side and line inductances $L_f$ and $L_g$ have opposite effects. This behavior differs from GFL or converters or SL-GFM converters with droop control, where stability depends only on the LCL resonant frequency, and $L_f$ and $L_g$ contribute in similar ways \cite{Parker2014,Wang2016, Liu2024a}.
\end{enumerate}

\begin{figure}[!t]
\centering
\includegraphics[width=\columnwidth]{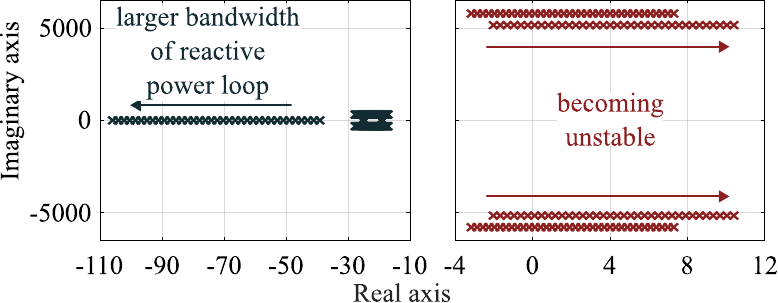}
\caption{Root loci when $k_q$ increases from 4 to 11 with corresponding bandwidth of RAP loop increasing from 6 Hz to 17 Hz.}
\label{fig_loci_kq}
\end{figure}

To further elaborate the aforementioned observations, a root locus analysis is performed. Fig. \ref{fig_loci_kq} shows the root locus diagram as $k_q$ increases from 4 to 11, where the RAP mode shifts from -39 to -106, corresponding to a bandwidth increasing from 6 Hz to 17 Hz. The resonance modes gradually move to the right-half plane (RHP), making the whole system unstable. This implies that LCL resonance limits the bandwidth of the RAP loop. The critical point at which the system becomes unstable is $k_q = 5.1$. If a larger $k_q$ is required for a quicker response, an additional AD strategy is required, which will be elaborated in Section \ref{sec_AD}.

\begin{figure}[!t]
\centering
\includegraphics[width=\columnwidth]{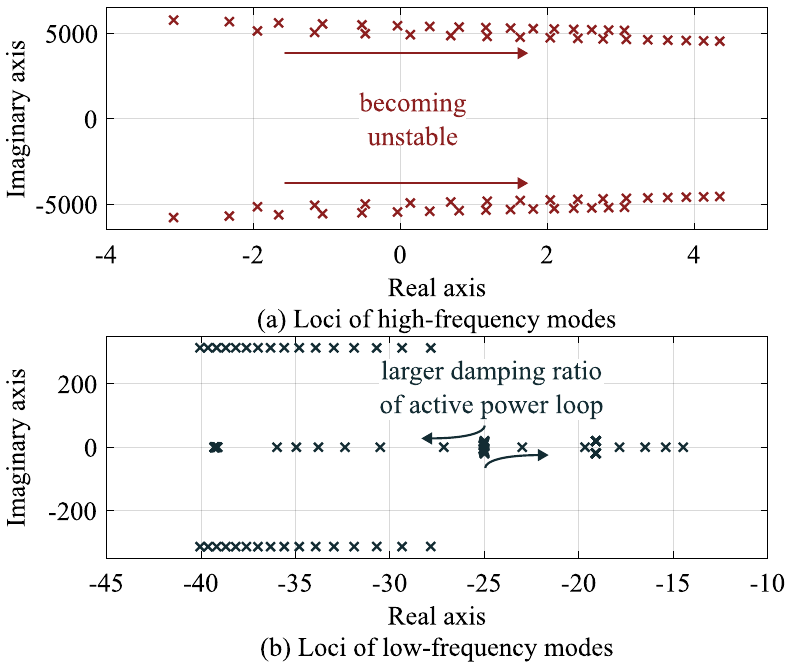}
\caption{Root loci when $L_g$ increases from 0.2 p.u. to 0.5 p.u.}
\label{fig_loci_lg}
\end{figure}

\begin{figure}[!t]
\centering
\includegraphics[width=\columnwidth]{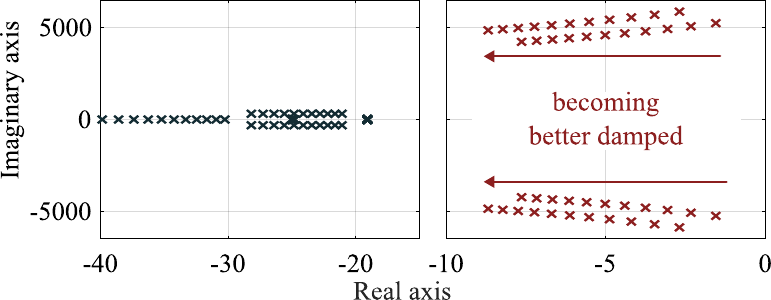}
\caption{Root loci when $L_f$ increases from 0.1 p.u. to 0.2 p.u.}
\label{fig_loci_lf}
\end{figure}

Fig. \ref{fig_loci_lg} and Fig. \ref{fig_loci_lf} show the root loci as $L_g$ increases from 0.2 p.u. to 0.5 p.u and $L_f$ increases from 0.1 p.u. to 0.2 p.u., respectively. As shown, a larger $L_g$, which implies a weaker grid, increases the damping ratio of the AP loop, aligning with the conventional understanding that GFM control performs better in weak grids \cite{Rosso2021}. However, the resonance modes move into the RHP, leading to system instability with undamped high-frequency oscillations. This demonstrates that LCL resonance fundamentally constrains the stable operation of SL-GFM converters in weak grids. Conversely, increasing $L_f$ enhances the stability of the resonance modes, pushing them further into the left-half plane (LHP). This indicates that $L_f$ and $L_g$ have opposite effects in SL-GFM converters, in contrast to their symmetrical role in GFL converters.

\begin{figure}[!t]
\centering
\includegraphics[width=\columnwidth]{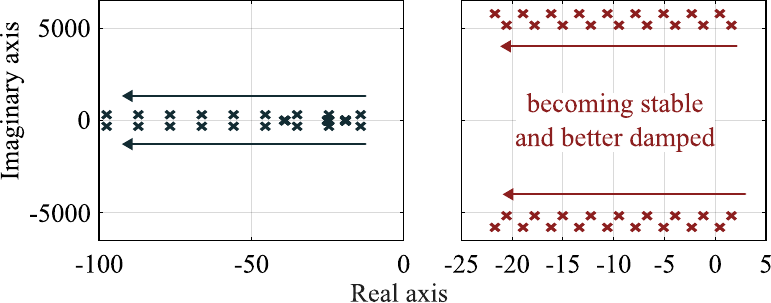}
\caption{Root loci when the ratio $X_g/R_g$ decreases from 10 to 2.}
\label{fig_LociRatio}
\end{figure}

Fig. \ref{fig_LociRatio} shows the root loci as the ratio $X_g/R_g$ decreases from 10 to 2, making the grid more resistive. As shown, the resonant modes are better damped in a more resistive grid, which is consistent with the common understanding on the relationship between resonance and resistive damping. In Section \ref{sec_AD}, we will propose an AD design method based on the equivalent resistor strategy. At the same time, the low-frequency power modes remain stable. Therefore, the case of $R_g = 0$ can be regarded as a worst-case scenario for the study of high-frequency resonance in the next section.

Moreover, the DC loop is influenced by the RAP loop through the coupling term as shown in Fig. \ref{fig_small_signal}. From Fig. \ref{fig_small_signal}, the relationship between $v_{dc}$ and $E_{rf}$ can be derived as:
\begin{align}
    \Delta v_{dc} = \frac{-(i_{d0}/v_{dc0})s}{(C_{dc}/\omega_n)s^2 + (k_{pdc} - E_{rf0}i_{d0}/v^2_{dc0})s + k_{idc}}\Delta E_{rf}
\end{align}
where $\Delta E_{rf}$ represents the influence of RAP loop on the DC-link voltage. As a result, a high-frequency unstable RAP loop, i.e., an unstable $\Delta E_{rf}$, is expected to destabilize the DC loop as well. Similarly, due to their coupling with the RAP loop, the variables in the AP loop and the WT are also unstable.

\subsection{Mechanism Analysis}

The preceding section demonstrated that the stability of high-frequency modes is highly sensitive to parameter variations and can significantly impact overall system stability, indicating that these modes cannot be neglected through simple frequency separation. This section further shows, from frequency response perspective, that such instability originates from a newly identified instability mechanism unique to SL-GFM systems with droop-I control. Specifically, it is shown that the RAP OL system undergoes a transition from minimum phase to nonminimum phase, due to the emergence of unstable OL poles, a phenomenon not observed in either GFL control or SL-GFM with droop control.

The findings in Section \ref{subsec_DetailedModel} suggest a need for further analysis of the RAP loop using the detailed model. In (\ref{eq_Gqe_sim}) and (\ref{eq_Gve_sim}), the TFs associated with the RAP loop in Fig. \ref{fig_small_signal} were derived based on the simplified static impedances. These TFs should now be reformulated using the detailed model given in (\ref{eq_did})-(\ref{eq_digq}). By assuming $R_g=0$, they can be expressed as:
\begin{align}
    \label{eq_GqE_detail}
    &G_{qE}(s) \coloneqq N_{qE}(s)/[(s^2 + \omega_n^2\omega_g^2)D_{LCL}(s)]\\
    \label{eq_GVE_detail}
    &G_{VE}(s) \coloneqq N_{VE}(s)/D_{LCL}(s)
\end{align}
where
\begin{align}
        &\begin{aligned}[b]
            N_{qE}(s) \coloneqq &\frac{\omega_n^2}{L_fL_gC_f}[v_{q0}(s^3 + \omega_n^2\omega_{res}^2s - 3\omega_n^2\omega_g^2s)\\
            &- v_{d0}(\omega_n^3\omega_g^3 - \omega_n^3\omega_{res}^2\omega_g - 3\omega_n\omega_gs^2)\\
            &- i_{gq0}L_g(s^2 + \omega_n^2\omega_g^2)(s^2 + \omega_n^2\omega_{res}^2 - \omega_n^2\omega_g^2)\\
            &-2i_{gd0}L_g\omega_n\omega_gs(s^2 + \omega_n^2\omega_g^2)]
        \end{aligned}\\
    \label{eq_Nve}
    &N_{VE}(s) \coloneqq \frac{\omega_n^2}{L_fC_fV_0}[v_{d0}(s^2 + \omega_n^2\omega_{res}^2 - \omega_n^2\omega_g^2) - 2v_{q0}\omega_n\omega_gs]\\
    \label{eq_Dlcl}
    &D_{LCL}(s) \coloneqq [s^2 + \omega_n^2(\omega_{res} + \omega_g)^2][s^2 + \omega_n^2(\omega_{res} - \omega_g)^2]
\end{align}
and $\omega_{res}$ is the LCL resonant angular frequency defined as \cite{Wang2016}:
\begin{equation}
\label{eq_wres}
    \omega_{res} \coloneqq \sqrt{(L_f + L_g)/(L_fL_gC_f)}
\end{equation}
which corresponds to 871 Hz using the parameters in Table \ref{tab_parameter}. 

\begin{figure}[!t]
\centering
\includegraphics[width=\columnwidth]{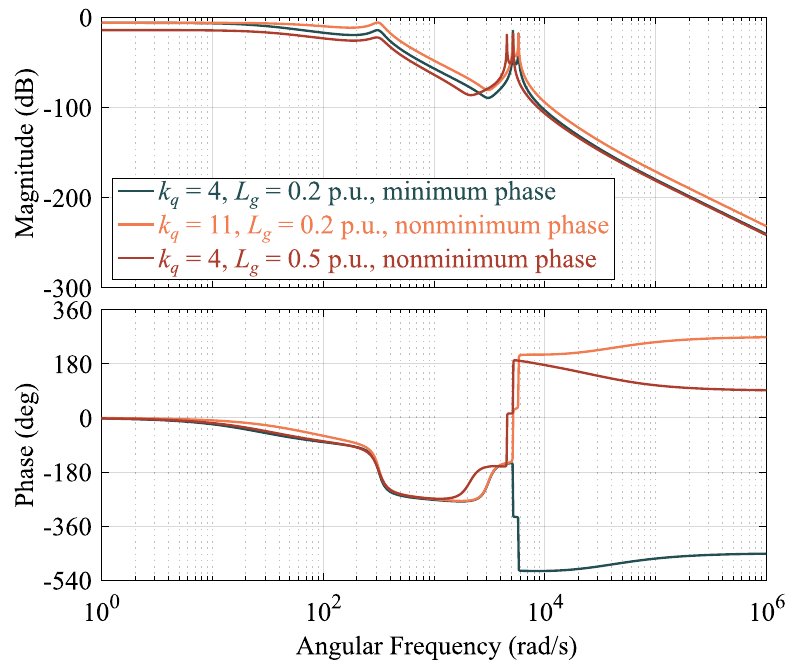}
\caption{Comparison of open-loop Bode plots of RAP loops.}
\label{fig_openbode}
\end{figure}

Finally, the OL TF of the RAP loop with the detailed model can be derived by substituting (\ref{eq_GqE_detail}) and (\ref{eq_GVE_detail}) into Fig. \ref{fig_small_signal} as
\begin{align}
\label{eq_Gq}
    G_q(s) \coloneqq &\frac{k_qN_{qE}(s)}{[sD_{LCL}(s) + k_qD_qN_{VE}(s)](s^2 + \omega_n^2\omega_g^2)}
\end{align}
where this OL TF will be used to evaluate the CL stability based on its frequency response. Its poles are calculated in Table \ref{tab_openpole} with different parameters. While the first case remains a minimum phase system, the other two cases exhibit RHP poles at resonance frequencies, indicating that the OL RAP control loop undergoes a transition from minimum phase to nonminimum phase as parameters vary. As shown in the OL Bode plots in Fig. \ref{fig_openbode}, all three cases satisfy the Bode's 0 dB gain and -$180^\circ$ phase crossover criteria. However, this observation highlights a critical limitation of applying previous conclusion on GFL \cite{Parker2014} or SL-GFM with droop control \cite{Liu2024a} to the SL-GFM with droop-I control, as using Bode plots alone to assess resonance stability without accounting for the presence of OL unstable poles is inadequate \cite{Franklin2019}. Although Fig. \ref{fig_openbode} suggests positive stability margins under crossover criteria, the latter two systems are in fact unstable, as confirmed by the root locus results in Fig. \ref{fig_loci_kq} and \ref{fig_loci_lg}, or when applying the complete Nyquist criterion with consideration of OL unstable poles. While the influence of nonminimum phase behavior on synchronous frequency resonance has been previously studied in \cite{Xiong2023}, this work demonstrates that LCL resonance is the dominant factor for SL-GFM converters with droop-I control. In contrast, prior studies on GFL and SL-GFM with droop control have only identified instability resulting from violations of the crossover criteria with insufficient stability margins, without observing a transition to nonminimum phase behavior. Consequently, SL-GFM with droop-I control exhibits inferior resonance stability characteristics.

\begin{table}[!t]
    \centering
    \renewcommand{\arraystretch}{1.3}
    \caption{Open-Loop Poles of RAP Loop with Detailed Model}
    \resizebox{\columnwidth}{!}{
    \begin{tabular}{ccc}
        \hline\hline\\[-3mm]
        Cases & High-frequency modes & Characteristics\\ \hline
        $k_q=4$ & $-2.9\pm j5786.2$&minimum phase\\ 
        $L_g=0.2$ p.u. & $-2.2\pm j5158.7$ &(stable)\\\hline
        $k_q=11$ & \color{chired}{$7.8\pm j5785.2$}&\\ 
        $L_g=0.2$ p.u. & \color{chired}{$9.9\pm j5159.9$} &\color{chired}{nonminimum phase}\\\cline{1-2}
        $k_q=4$ & \color{chired}{$3.2\pm j5176$} &\color{chired}{(unstable)}\\
        $L_g=0.5$ p.u. &\color{chired}{$4.3\pm j4548.8$}\\[1.4ex]
        \hline\hline
    \end{tabular}
    }
    \label{tab_openpole}
\end{table}

As shown in Fig. \ref{fig_small_signal}, the droop-I control introduces feedback paths involving both $q$ and $V$, which form an inner closed loop composed of $D_q$, $k_q/s$, and $G_{VE}(s)$. Since $k_q$ is related to the RAP control and $G_{VE}(s)$ reflects the LCL filter dynamics, the inner loop effectively couples the low-frequency RAP control with the high-frequency LCL dynamics. The analysis implies that the instability originates from the RAP control loop rather than from the DC-side dynamics. When different DC input sources are taken into account, the analysis and following AD design based on the RAP control loop will remain valid and effective.

Furthermore, as observed from (\ref{eq_Dlcl}) and (\ref{eq_wres}), $L_f$ and $L_g$ appear symmetrically and have identical effects on $\omega_{res}$ and $D_{LCL}(s)$. However, $N_{VE}(s)$, from (\ref{eq_Nve}), depends on $L_f$ through the term $\frac{\omega_n^2}{L_fC_fV_0}$. Therefore, for a same resonant frequency, a larger $L_f$ results in a smaller $N_{VE}(s)$ and thus a reduced influence of the inner feedback branch. On the contrary, a larger $L_g$ is effectively equivalent to a smaller $L_f$, leading to a stronger influence of the inner feedback branch.

\begin{figure}[!t]
\centering
\includegraphics[width=\columnwidth]{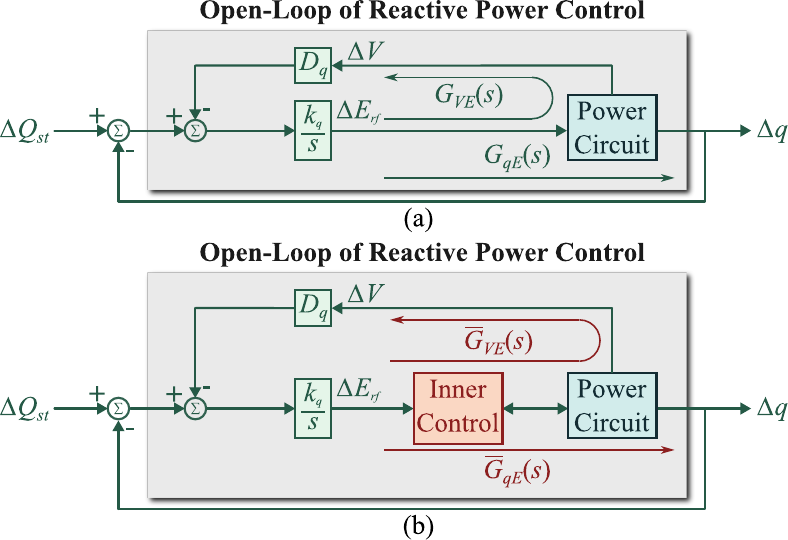}
\caption{Schematic diagram of loops and corresponding transfer functions of RAP loop. (a) Without inner control. (b) With inner control.}
\label{fig_AnalysisFramework}
\end{figure}

The analysis framework can include the inner control as well. An equivalent schematic of the loops and their corresponding TFs of the RAP loop in Fig. \ref{fig_small_signal} can be shown in Fig. \ref{fig_AnalysisFramework}(a), where $G_{VE}(s)$ in the inner loop depends only on the power circuit. When the inner control is included, the corresponding schematic can be shown in Fig. \ref{fig_AnalysisFramework}(b). In this context, nonminimum phase behavior can generally be eliminated with the additional degrees of freedom provided by the inner control.

It is worth emphasizing that the OL RHP poles are not the only cause of resonant oscillations. Even for an OL minimum phase, resonant oscillations may still arise if the -180$^\circ$ phase and 0 dB gain crossover criterion is violated. This is the mechanism of conventional resonance, which has been well studied in literature \cite{Liu2024a}. Although the two mechanism are different, they both lead to unstable CL poles at the resonant frequencies, resulting in resonant oscillations.

\subsection{Comment on Different $q-V$ Control Strategies}

\begin{figure}[!t]
\centering
\includegraphics[width=\columnwidth]{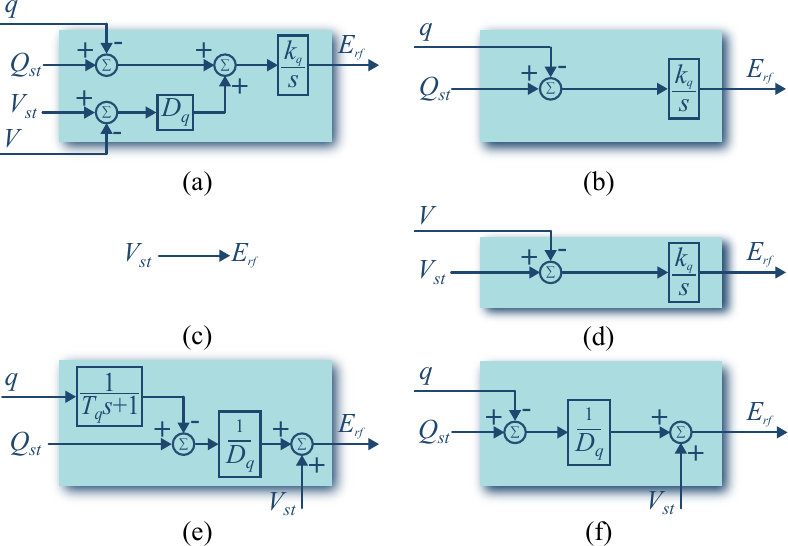}
\caption{Typical RAP control structures illustrated for comparison. (a) Droop-I control; (b) RAP control; (c) Fixed voltage control; (d) Voltage control; (e) Droop control; (f) Pure droop control.}
\label{fig_avr}
\end{figure}

This section compares the resonance behavior of SL-GFM converters under six different RAP controls, given the significant influence of RAP control on resonance modes. Fig. \ref{fig_avr} summarizes the considered control structures: (a) droop-I control, used in the previous sections and as a reference for the comparison; (b) RAP control (without droop characteristics)\cite{Oraa2024}; (c) fixed voltage control; (d) voltage control; (e) droop control \cite{Liu2022}; and (f) pure droop control (without a power filter) \cite{Pan2020}. Meanwhile, for a fair comparison, all corresponding RAP modes are adjusted to about -40, consistent with Fig. \ref{fig_pole}. Only the RAP mode and resonance modes are shown in this Section.

\subsubsection{RAP Control} 
\begin{figure}[!t]
\centering
\includegraphics[width=\columnwidth]{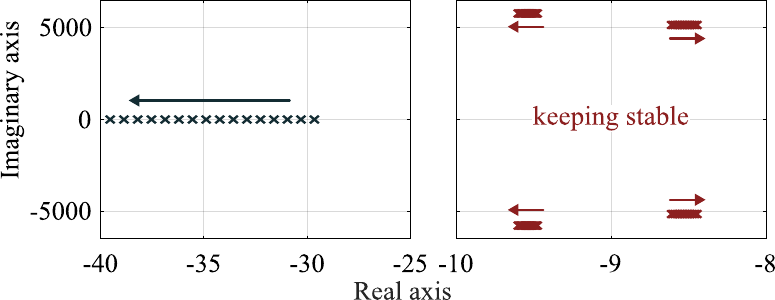}
\caption{Root loci of SL-GFM converter with RAP control of Fig. \ref{fig_avr}(b) when $k_q$ increases.}
\label{fig_loci_kq_q}
\end{figure}

Its mode can be adjusted via the integral gain, with the corresponding root loci shown in Fig. \ref{fig_loci_kq_q}. The results indicate that changes in RAP bandwidth have minimal impact on the resonance modes. When the RAP mode is positioned at -40, the resonance modes shift further away from the imaginary axis compared to Fig. \ref{fig_pole}, suggesting improved system stability.

\subsubsection{Fixed Voltage Control} The inverter voltage is directly assigned without regulation, equivalent to Fig. \ref{fig_avr}(b) with $k_q=0$. Referring to Fig. \ref{fig_loci_kq_q} and considering a gradual reduction of $k_q$ to zero, the critical resonance mode will move to the left. This indicates improved stability compared to RAP control.

\subsubsection{Voltage Control} 
\begin{figure}[!t]
\centering
\includegraphics[width=\columnwidth]{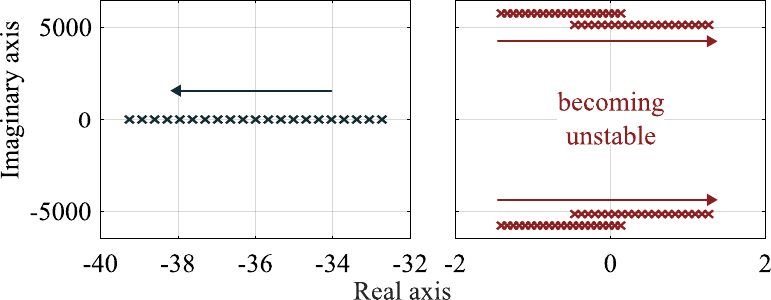}
\caption{Root loci of SL-GFM converter with voltage control of Fig. \ref{fig_avr}(d) when $k_q$ increases.}
\label{fig_loci_kq_V}
\end{figure}

Its mode can also be tuned via the integral gain, with the corresponding root loci presented in Fig. \ref{fig_loci_kq_V}. The results indicate that increasing voltage regulation speed adversely affects the stability of the resonance modes. As the low frequency mode moves to -40, the resonance modes shift into the RHP, rendering the system unstable. This suggests that its resonance stability is worse than that of the droop-I control.

\subsubsection{Droop Control} 
\begin{figure}[!t]
\centering
\includegraphics[width=\columnwidth]{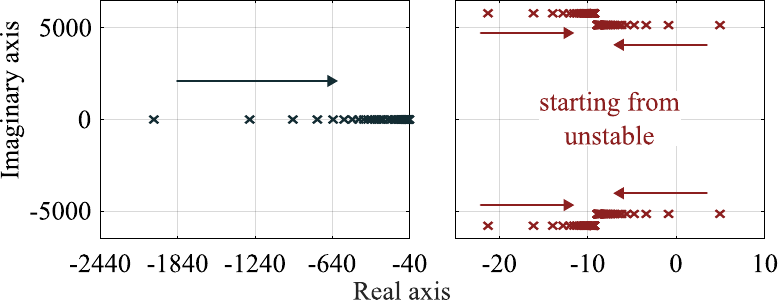}
\caption{Root loci of SL-GFM converter with droop control of Fig. \ref{fig_avr}(e) when $T_q$ increases.}
\label{fig_loci_Tq}
\end{figure}

The steady-state droop characteristic is expressed as $\Delta E_{rf} = -1/D_q\Delta q$, where the associated voltage $E_{rf}$ is the inverter output voltage rather than $V$. As a result, the actual droop characteristic between $q$ and $V$ is affected by the voltage drop across the filter elements $L_f$ and $C_f$ as well as additional blocks such as virtual impedance. Its mode can be adjusted via the time constant of the RAP filter $T_q$, with the corresponding root loci shown in Fig. \ref{fig_loci_Tq}. The results indicate that $T_q$ has opposite effects on the two resonance modes, with the critical mode gradually shifting from the RHP to the LHP, improving system stability. When the RAP mode reaches -40, the critical resonance mode is positioned farthest in the LHP compared to all previously discussed strategies, suggesting that this approach provides the best stability properties for the resonance modes.

\subsubsection{Pure Droop Control} The RAP $q$ is directly fed back to the droop mechanism without a pre-filter, which corresponds to Fig. \ref{fig_avr}(e) with $T_q=0$. As observed in Fig. \ref{fig_loci_Tq}, a small $T_q$ results in unstable resonance modes. Therefore, pure droop control cannot be directly applied to SL-GFM converters unless the resonance effects are properly mitigated. It is worth noting that existing research on SL-GFM resonance is based on the pure droop control \cite{Liu2024a}, which attributes the observed instability to insufficient stability margins caused by undesired crossover, which AD used to suppress such crossover effects. As discussed in comment 4) regarding droop control, $T_q$ contributes to enhancing resonance stability and droop-I control exhibits a higher risk to resonance instability as transition to nonminimum phase may occur.

In summary, the resonance stability of various RAP control structures in Fig. \ref{fig_avr} can be ranked as
\begin{equation}   \underbrace{\text{(e)}>\text{(c)}>\text{(b)}>\text{(a)}>\text{(d)}}_\text{Stable or could be tuned to be stable}>\underbrace{\text{(f)}}_\text{Unstable}\notag
\end{equation}

\section{Active Damping Strategy}\label{sec_AD}

From the perspective of the CL frequency characteristics, the resonance implies that the system contains resonant peaks around the resonant frequencies. The resonance not only influences the high-frequency stability of the system but also deteriorates the power quality by amplifying disturbances/noises around these resonant frequencies. Therefore, proper AD designs are required to provide sufficient damping.

\subsection{Selection of Active Damping Implementation}

\begin{figure}[!t]
\centering
\includegraphics[width=0.52\columnwidth]{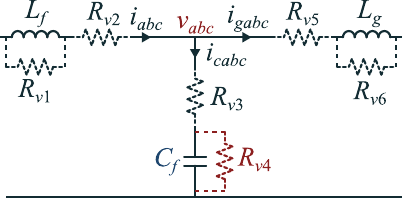}
\caption{Equivalent virtual resistors of various AD strategies.}
\label{fig_virtual_resistor}
\end{figure}

Ideally, AD introduces a virtual resistor in one of six possible locations, labeled $R_{v1}$ to $R_{v6}$ in Fig. \ref{fig_virtual_resistor}. As discussed in \cite{Liu2020}, all virtual resistors can be implemented using feedback from $i_{abc}$, $i_{gabc}$, $i_{cabc}$, or $v_{abc}$, leading to 24 possible AD configurations. Although the AD strategies based on hybrid feedback variables may improve the robustness by offering additional control degrees of freedom, it inevitably increases the complexity of both the structure and the parameters tuning \cite{Pooja2024}. Therefore, when resonance can be effectively suppressed, the AD strategy usually uses the single-variable feedback scheme. For GFL converters, $i_{gabc}$ feedback to implement $R_{v4}$ has been identified as the most suitable approach in terms of number of sensors, implementation complexity, and damping effectiveness. However, applying these same selection criteria to SL-GFM converters leads to a different optimal choice, as will be discussed in this section.

\subsubsection{Number of Sensors} For SL-GFM converters, both $i_{gabc}$ and $v_{abc}$ are measured for power calculation. Therefore, the AD should be implemented using only these variables whenever possible, avoiding additional sensors. This requirement leads to 12 possible implementations.

\subsubsection{Complexity} Using $i_{gabc}$ results in controllers where the order of the numerator is higher by at least two than that of the denominator. The same applies when constructing $R_{v1}$, $R_{v3}$, or $R_{v6}$ with $v_{abc}$ \cite{Liu2020}. These choices are not favorable in practice due to implementation complexity. Consequently, only three viable options remain: using $v_{abc}$ to construct $R_{v2}$, $R_{v4}$, or $R_{v5}$. Notably, while $i_{gabc}$ is preferred in GFL converters, it is recognized as not being the most suitable approach for SL-GFM converters at this stage.

\subsubsection{Effectiveness} The added AD should minimize its negative impact on the original system. In GFL converters, the primary concern is whether the damped resonance limits the current loop bandwidth. A key issue in GFM converters is the steady-state power regulation of the AP and RAP loops. This paper evaluates the steady-state coupling strength by analyzing the steady-state change in RAP caused by an AP step, defined as $K_{qp} \coloneqq \lim_{s\rightarrow0}\frac{\Delta q(s)}{\Delta P_{st}(s)}$. Fig. \ref{fig_virtual_resistor} indicates that a better damped resonance mode leads to stronger steady-state coupling when using $R_{v5}$. In contrast, $R_{v2}$ and $R_{v4}$ do not affect the steady-state power regulation due to the robust droop characteristics provided by droop-I control \cite{Chen2021}, making them more favorable. 


As a result, implementing capacitor voltage feedback to construct $R_{v2}$ or $R_{v4}$ is the preferred approach for SL-GFM converters. The corresponding AD controllers are defined as:
\begin{equation}
\label{eq_Gad}
    G_{ad}(s) \coloneqq \begin{cases}
        k_ds + k_i/s & \rm{for}~\it{R_{v\text{2}}}\\
        k_ds & \rm{for}~\it{R_{v\text{4}}}
    \end{cases}
\end{equation}
where $k_d$ and $k_i$ are parameters to be designed, 
and $R_{v4}$ is chosen so as to have a simple implementation in practice. It is noted that a similar AD strategy based on $v_{abc}$ has been previously applied in GFL converters \cite{Dannehl2010}. Nevertheless, its benefits and design in the context of SL-GFM converters has not been reported or analyzed in the literature.

\subsection{Design of Active Damping}

\begin{figure}[!t]
\centering
\includegraphics[width=\columnwidth]{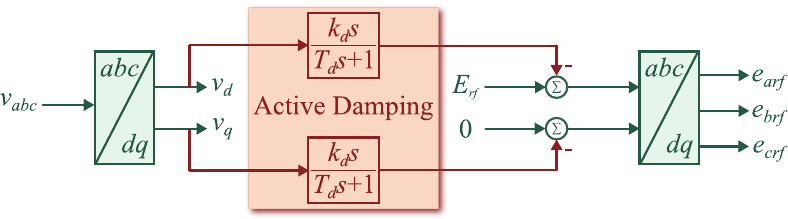}
\caption{Block diagram of AD strategy for SL-GFM converters.}
\label{fig_ActiveDamping}
\end{figure}

Based on (\ref{eq_Gad}), the AD is implemented in practice as:
\begin{equation}
\label{eq_Gad_im}
    G_{ad}(s) \coloneqq k_ds/(T_ds + 1)
\end{equation}
where $T_d$ is a time constant. Finally, the implemented AD strategy in the $dq$ reference frame is shown in Fig. \ref{fig_ActiveDamping}.

\begin{figure}[!t]
\centering
\includegraphics[width=\columnwidth]{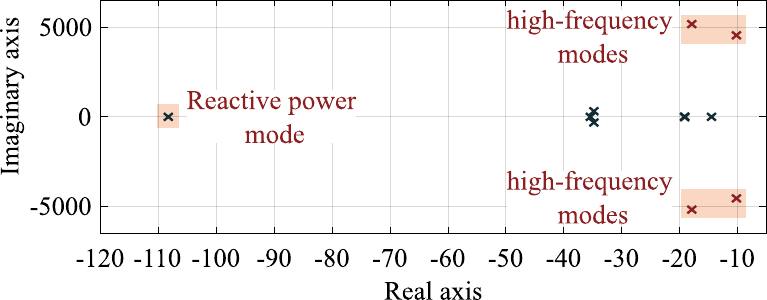}
\caption{Designed eigenvalues based on worst-case scenario.}
\label{fig_DampingPole}
\end{figure}

Based on the analysis in the previous sections, a step-by-step design procedure for $G_{ad}(s)$ is proposed as follows:
\begin{itemize}
    \item \textit{Step 1}: Determine specifications in terms of the placement of the RAP mode and $L_g$. As shown in Section \ref{subsec_DetailedModel}, a RAP mode farther from the imaginary axis and a larger $L_g$ will deteriorate the resonant stability. Therefore, we can just focus on a worst-case scenario, where the RAP mode is placed at the farthest distance and $L_g$ takes its largest value. Under normal operating conditions, the system is expected to operate away from the extreme case, therefore ensuring robustness for all actual RAP modes and $L_g$ values within the specified maximum limits. This approach is illustrated in this paper by requiring the resonance to be damped when the RAP mode is placed at about -110, corresponding to 17 Hz, with $L_g$ set to 0.5 p.u., which corresponds to a SCR of 2.
    \item \textit{Step 2}: Determine the stability margin in terms of the resonance modes. $k_d$ should make the resonant modes in the LHP for stable operation. To guarantee a quick damping and higher robustness, it is favorable that the high resonant modes place not only in the left-half plane but also away from the imaginary axis, i.e., a certain margin. This ensures that even under the worst-case scenario, the system is still away from the critical boundary. This paper, as an example, requires the real parts to be less than -10 to ensure a sufficient stability margin.
    \item \textit{Step 3}: Determine $k_d$ via small-signal analysis. This step assumes $T_d=0$. With the designed AD strategy of $G_{ad}(s) = k_ds$, the inverter voltage are defined by
    \begin{align}
        &e_d = E_{rf} - k_dsv_d\\
        &e_q = -k_dsv_q    
    \end{align}
    which, combining with equations in Section \ref{sec_HF}, compose the mathematical model of the system with proposed active strategy. A good selection of $k_d$ should suppress the resonance with different $k_q$ and $L_g$. A small-signal analysis of root locus with respect to variations of $k_d$ can be performed under the worst-case scenario defined in $Step~1$. Then $k_d$ is selected such that the resonant modes are placed away from the imaginary axis with the distance margin defined in $Step~2$. The parameters used result in $k_d=3.3\times10^{-6}$. The system eigenvalues, shown in Fig. \ref{fig_DampingPole}, confirm that the placement meets the defined specifications and stability margin. Notably, the small-signal analysis is conducted under the worst-case scenario without using an exact line impedance value. Since line impedance does not typically vary continuously in practice, the AD can also combine with impedance estimation by constructing a look-up table mapping $L_g$ to the corresponding $k_d$. The gain $k_d$ can then be online updated based on the estimated value of $L_g$ to delicately tune the performance. Furthermore, in table \ref{tab_parameter}, $D_q = 10$ is used an example. For different values of $D_q$ set by the higher-level supervisor, the corresponding AD parameters can be rederived accordingly. This design procedure ensures that the proposed AD does not adversely affect higher-level settings.
    \item \textit{Step 4}: Determine $T_d$. $T_d$ is introduced to avoid pure derivative action without significantly affecting the desired transfer function characteristics at the resonant frequency. It is set to $T_d=8\times10^{-5}$, ensuring a break-point frequency of 2 kHz, which is twice the designed LCL resonant frequency. Finally, $G_{ad}(s)$ is designed as:
    \begin{equation} 
    G_{ad}(s) = 3.3\times10^{-6}s/(8\times10^{-5}s + 1)
    \end{equation}
    \item \textit{Step 5}: Validation. The effectiveness of $G_{ad}(s)$ can be validated via simulation/experiments by testing various combinations of $k_q$ and $L_g$ within the predefined ranges in \textit{Step 1} to verify that the system can remain stable without triggering high-frequency resonance.
\end{itemize}

It is worth noting that the designed controller in (\ref{eq_Gad_im}) behaves as a high-pass filter, and therefore primarily acts on high frequency dynamics. In contrast, for low-frequency phenomena in SL-GFM converters, the controller has negligible influence on the dynamics.

\section{Comparison of LCL Resonance Between GFL and SL-GFM Converters}\label{sec_comparison}

LCL resonance and its damping strategies have been extensively studied in GFL converters. Therefore, this section summarizes the findings in the previous sections and provides a comparison between SL-GFM (with droop-I RAP control) and GFL converters, highlighting their distinct characteristics in relation to LCL resonance and damping. The key features are compared in Table \ref{tab_comparison}.

\begin{table}[!t]
    \setlength{\tabcolsep}{3.5pt}
    \renewcommand{\arraystretch}{1.3}
    \centering
    \caption{Comparison of LCL Resonance Between SL-GFM and GFL Converters}
    \resizebox{\columnwidth}{!}{
    \begin{tabular}{cccc}
        \hline\hline\\[-3mm]
        \multicolumn{2}{c}{} & SL-GFM Converter & GFL Converter \\ \hline
        \multirow{2}{*}{\begin{tabular}[c]{@{}c@{}}Power\vspace{-1mm}\\stage\end{tabular}} & Critical parameters &  $L_f$, $C_f$, $L_g$ & $L_f$, $C_f$, $L_g$\\ \cline{2-4}
        & Roles of $L_f$ and $L_g$& {\color{red}\textbf{Opposite}}&  {\color{red}\textbf{Identical}}\\ \hline
        \multirow{6}{*}{Control} &Measured variables&\begin{tabular}[c]{@{}c@{}} Grid currents\vspace{-1mm}\\{\color{red}\textbf{Capacitor voltages}}\end{tabular}&Grid currents\\\cline{2-4}
         & Critical loops &  {\color{red}\textbf{RAP control}} & {\color{red}\textbf{Current control}}\\ \cline{2-4}
         & Open-loop system&\begin{tabular}[c]{@{}c@{}} {\color{red}\textbf{May become}}\vspace{-1mm}\\{\color{red}\textbf{Nonminimum phase}}\end{tabular}& {\color{red}\textbf{Minimum phase}}\\\cline{2-4}
         & Stability Analysis&\begin{tabular}[c]{@{}c@{}}Root loci\vspace{-1mm}\\Nyquist criterion\end{tabular}&\begin{tabular}[c]{@{}c@{}}Root loci\vspace{-1mm}\\Nyquist criterion\vspace{-1mm}\\{\color{red}\textbf{Bode plot}}\end{tabular}\\\hline
         \multirow{3.5}{*}{\begin{tabular}[c]{@{}c@{}}Selected\vspace{-1mm}\\AD\end{tabular}}
         & Selection criteria&\begin{tabular}[c]{@{}c@{}}number of sensors\vspace{-1mm}\\Complexity\vspace{-1mm}\\{\color{red}\textbf{Bandwidth}}\end{tabular}&\begin{tabular}[c]{@{}c@{}}number of sensors\vspace{-1mm}\\Complexity\vspace{-1mm}\\{\color{red}\textbf{Coupling strength}}\end{tabular}\\\cline{2-4}
         & Virtual resistor & Paralleled with $C_f$  & Paralleled with $C_f$\\ \cline{2-4}
         & Feedback variables&{\color{red}\textbf{Capacitor voltages}}& {\color{red}\textbf{Line currents}}\\[1.4ex]
        \hline\hline
    \end{tabular}
    }
    \label{tab_comparison}
\end{table}

\begin{itemize}
    \item \textit{Power Stage}: In both GFL and SL-GFM converters, LCL filter parameters naturally affect resonance. However, in GFL converters, the resonant frequency is the primary concern, with $L_f$ and $L_g$ having identical effects \cite{Parker2014,Wang2016}. In contrast, $L_f$ and $L_g$ have opposite impact on the resonance modes in SL-GFM converters.
    \item \textit{Control}: In GFL converters, resonance stability primarily depends on current control and can be assessed using root loci for the CL system or frequency responses (Nyquist and Bode plots) for OL minimum phase systems. In contrast, the critical loop in SL-GFM converters is the RAP control. Since the SL-GFM's OL system may become nonminimum phase/unstable, the Bode plot for evaluating stability margins needs to be carefully used in conjunction with any OL unstable poles present. Additionally, SL-GFM control uses both grid current and capacitor voltage for power calculation, enabling more flexible AD implementations without extra sensors than GFL converters, which rely only on grid current.
    \item \textit{Selected AD}: AD for both GFL and SL-GFM converters can be selected using similar criteria. However, in GFL converters, AD effectiveness should not compromise control bandwidth whenever possible, whereas in SL-GFM converters, minimizing the coupling between AP and RAP loops is more critical. As results from the AD selection criteria discussed, line current feedback is preferred for GFL converters \cite{Wang2016,Liu2020}, while capacitor voltage feedback is more suitable for SL-GFM converters. Despite this difference, the optimal placement of the virtual resistor in both cases is recognized as being parallel to the filter capacitor.
\end{itemize}

\section{Experimental Tests}\label{sec_test}

\begin{figure}[!t]
\centering
\includegraphics[width=\columnwidth]{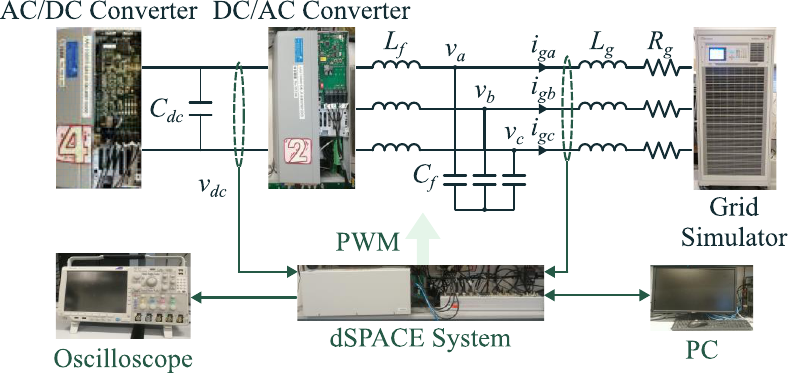}
\caption{Experimental setup.}
\label{fig_Setup}
\end{figure}

This section demonstrates that instability originates from LCL resonance, thus verifying the analysis and the effectiveness of the proposed AD. The experimental setup is associated with a DS1007 dSPACE system, as shown in Fig. \ref{fig_Setup}. The parameters are downscaled to 3kW/200V according to Table \ref{tab_parameter}, unless otherwise specified.

For comparison, the current-feedback-based AD \cite{Liu2024a} via $2\times10^{-5}s/(3.5\times10^{-3}s+1)$ used for SL-GFM with droop control is also tested as well. Although the chosen values satisfy the crossover criteria with enough margins, the results show that they fail to provide sufficient damping to suppress the emergence of nonminimum phase behavior and eliminate high-frequency resonances with droop-I control. Regardless of new current sensors, this outcome highlights the significance of the newly identified instability mechanism and the need to explicitly consider the nonminimum phase behavior during AD design to droop-I based SL-GFM systems. The key observations are summarized in Table \ref{tab_SettlingTime} for a comparison, where the frequencies listed in the "Stability" column correspond to the dominant frequency components with the largest magnitude obtained via FFT analysis of the measured waveforms.

\begin{table*}[!t]
    \renewcommand{\arraystretch}{1.3}
    \centering
    \setlength{\tabcolsep}{5pt}
    \caption{Summary of Experimental Results}
    \resizebox{\textwidth}{!}{
    \begin{tabular}{ccccccc}
        \hline\hline\\[-3mm]
        \multicolumn{3}{c}{Cases} & Resonant gain margin (dB) & Phase & Stability & Settling time (s)\\\hline
        & $k_q = 4$&no AD&21.3&minimum phase & stable & -\\\cline{2-7} 
         $L_g = 0.2$ p.u.& \multirow{3}{*}{$k_q = 14$} &no AD&26.1&nonminimum phase & resonant (820 Hz) & $\infty$\\
         $f_{res} = 871$ Hz& &Current-based AD \cite{Liu2024a}&20.7&nonminimum phase & resonant (820 Hz) & $\infty$\\ 
         & &Proposed AD&24.7&minimum phase & stable & \textless~0.5\\\hline
         & $k_q = 2$&no AD&5.32&minimum phase & stable & -\\\cline{2-7} 
         $L_g = 0.5$ p.u.& \multirow{3}{*}{$k_q = 5$} &no AD&33.8&nonminimum phase & resonant (720 Hz) & $\infty$\\
         $f_{res} = 774$ Hz& &Current-based AD \cite{Liu2024a}&10.1&minimum phase & stable & \textgreater~1\\ 
         & &Proposed AD&44&minimum phase & stable & \textless~0.5\\[1.4ex]
        \hline\hline
    \end{tabular}
    }
    \label{tab_SettlingTime}
\end{table*}

\begin{figure}[!t]
\centering
\includegraphics[width=\columnwidth]{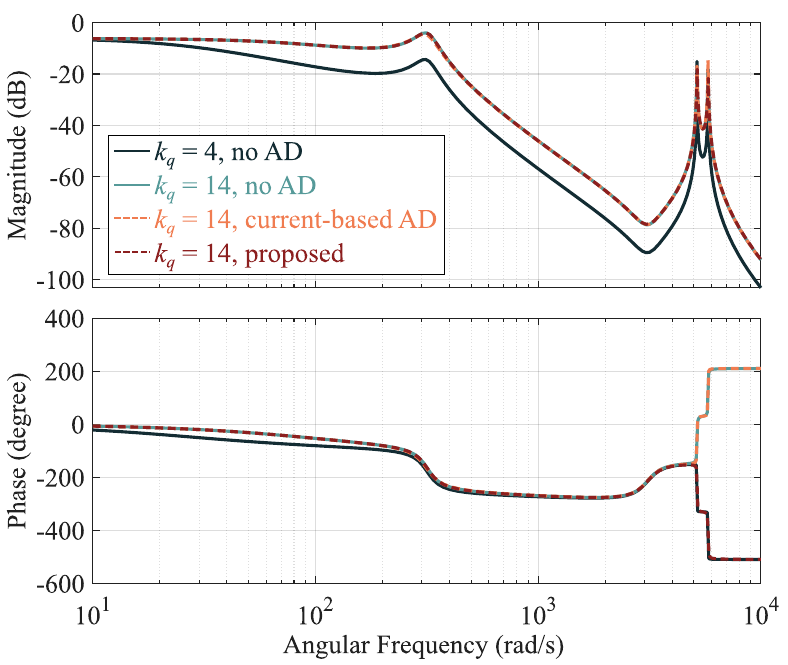}
\caption{Comparison of OL Bode plots of RAP loops with parameters in Table \ref{tab_parameter}.}
\label{fig_ExperimentalBode1}
\end{figure}

\begin{figure}[!t]
\centering
\includegraphics[width=\columnwidth]{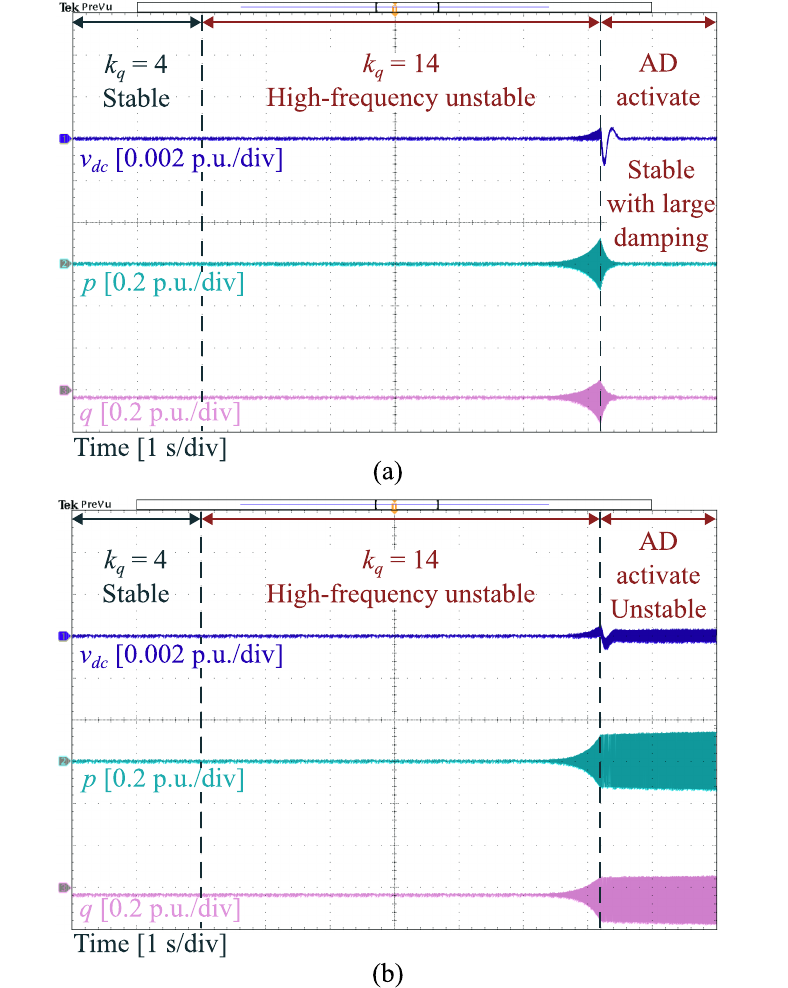}
\caption{Experimental results with parameters in Table \ref{tab_parameter}. (a) Voltage-feedback-based AD design considering nonminimum phase behavior. (b) Current-feedback-based AD design in droop control using crossover criterion without considering nonminimum phase behavior.}
\label{fig_Experimental_kq}
\end{figure}

Using the parameters in Table \ref{tab_parameter}, the OL Bode plots of the RAP loops and the experimental results of the DC-link voltage $v_{dc}$, output AP $p$, and output RAP $q$ are compared in Fig. \ref{fig_ExperimentalBode1} and Fig. \ref{fig_Experimental_kq}, respectively. Initially, the system operates stably without resonance, as it is minimum phase with a positive gain margin of 21.3 dB around the resonant frequency. At $t=2$ s, $k_q$ is increased from 4 to 14 to raise the bandwidth of the RAP mode from 6 Hz. It is noted that instability is observed with growing high-frequency oscillations in all three loops. Fig. \ref{fig_Experimental_kq} demonstrates that the instability originates from the LCL resonance rather than low-frequency interactions. Meanwhile, RAP control will influence the resonance stability, where a larger RAP control bandwidth deteriorates stability. Although the system still has a positive gain margin of 26.1 dB near the resonant frequency, this does not guarantee stability because the resonance originates from nonminimum phase behavior. Once the proposed AD is activated, the system is driven back to minimum phase with a gain margin of 24.7 dB, successfully attenuating the oscillations within 0.5 s. In contrast, the droop control-based AD, while capable of slowing down the rate of divergence, fails to effectively suppress the high-frequency oscillations. As shown in Table \ref{tab_SettlingTime}, the system remains nonminimum phase, and therefore its positive gain margin of 20.7 dB does not imply resonant stability. Meanwhile, Fig.\ref{fig_Experimental_kq}(a) shows the steady-state active and reactive power values remain unchanged before and after the activation of the proposed AD strategy and confirms that it does not affect steady-state power regulation characteristics.

\begin{figure}[!t]
\centering
\includegraphics[width=\columnwidth]{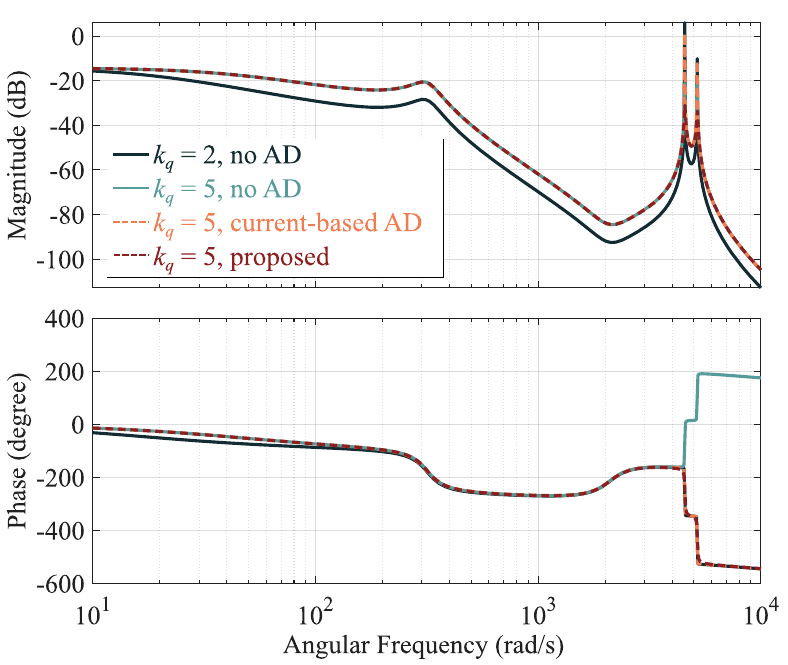}
\caption{Comparison of OL Bode plots of RAP loops with $L_g = 0.5$ p.u.}
\label{fig_ExperimentalBode2}
\end{figure}

\begin{figure}[!t]
\centering
\includegraphics[width=\columnwidth]{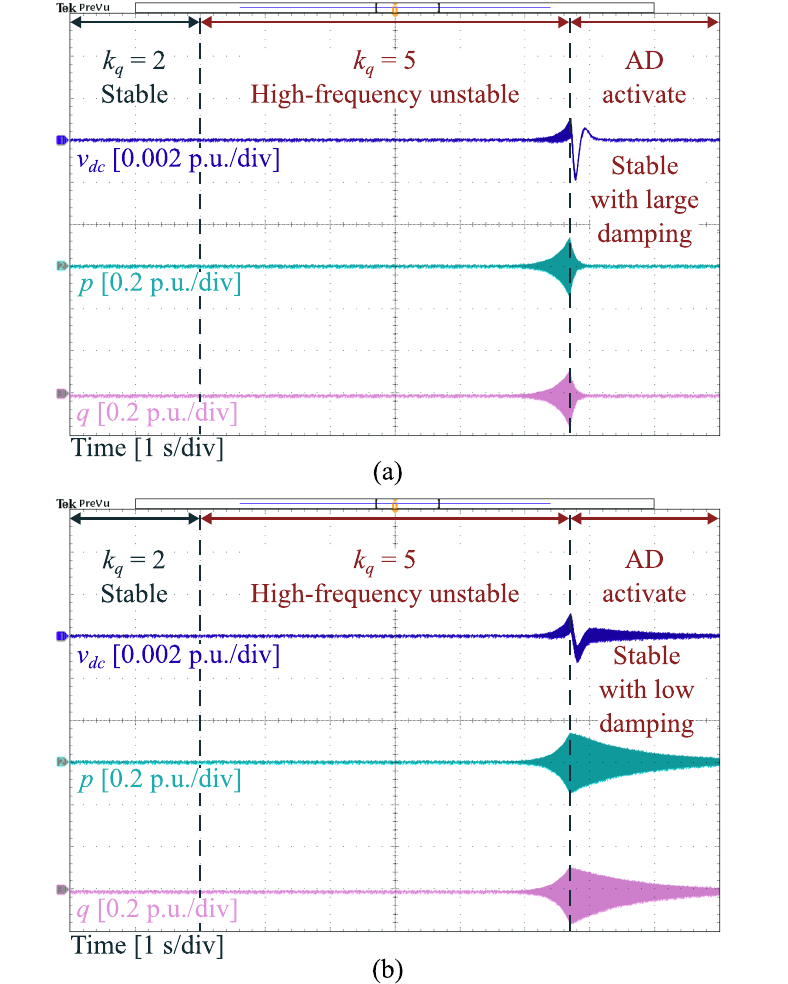}
\caption{Experimental results with $L_g=0.5$ p.u. (a) Voltage-feedback-based AD design considering nonminimum phase behavior. (b) Current-feedback-based AD design in droop control using crossover criterion without considering nonminimum phase behavior.}
\label{fig_ExperimentalLg}
\end{figure}

Another case is tested with $L_g$ increased to 0.5 p.u., where $f_{res}$ turns to be 774 Hz and the results are shown in Fig. \ref{fig_ExperimentalBode2} and Fig. \ref{fig_ExperimentalLg}. Due to the larger $L_g$, $k_q$ is set to a smaller value of 2 to ensure stable initial operation, for which the system is minimum phase with a gain margin of 5.32 dB. When $k_q$ is increased to 5, the OL system becomes nonminimum phase, despite having a larger gain margin of 33.8 dB. Compared to the case with $L_g=0.2$ p.u. in Fig. \ref{fig_Experimental_kq}, the system still exhibits pronounced high-frequency oscillations despite a smaller $k_q$ being used. This confirms that a larger $L_g$, indicative of a weaker grid, deteriorates the resonance stability of the SL-GFM converter. Similar to Fig. \ref{fig_Experimental_kq}, the proposed AD successfully suppresses the oscillations within 0.5 s by driving the OL system to minimum phase with a large positive gain margin of 44 dB. Although the droop-based AD can also make the system minimum phase and stabilize the oscillations in this case, it results in a small gain margin of 10.1 dB. Therefore, its dynamic response is slow and exhibits low damping, with the resonance requiring more than 1 s to be fully mitigated. Both Fig. \ref{fig_Experimental_kq} and \ref{fig_ExperimentalLg} demonstrate that, for AD design in droop-I control, robustness against variations in RAP bandwidth and grid strength requires explicit consideration of nonminimum phase poles. Otherwise, the high-frequency oscillations may persist, as seen with AD originally designed for droop control.

\section{Conclusion}\label{sec_con}
This paper identifies a novel high-frequency instability mechanism in SL-GFM converters with droop-I control. The following conclusions 
can be drawn:
\begin{enumerate}
    \item The SL-GFM converter with droop-I control may become an open-loop nonminimum phase system with unstable poles associated with LCL resonance, which is not observed neither in conventional GFL nor in SL-GFM with droop control. As a result, LCL resonance can significantly impact the system CL stability and cannot be simply ignored, even when the resonant frequency is much higher than that of the power loops.
    \item The stability properties associated with the LCL resonance are not decoupled with the low-frequency power loops, with RAP-voltage control playing an important role. A faster RAP loop usually worsens the stability properties of resonance modes. Notably, compared to droop control, the droop-I control has a reduced robustness with more severe risk of high-frequency instability.
    \item Unlike with GFL converters, where the primary concern is the LCL resonant frequency, SL-GFM with droop-I control exhibits distinct resonance stability characteristics. Although the inverter-side and grid-side inductors equally determine the LCL resonant frequency, they play different roles for the stability properties of the resonance modes  of the SL-GFM converter. Notably, a larger $L_g$ deteriorates such stability characteristics, constraining the stable operation of the SL-GFM converter in weak grids.
    \item In contrast to GFL converters, which typically employ AD via current feedback, SL-GFM converters have improved stability properties when using AD based on capacitor voltage feedback. A worst-case scenario can be defined to facilitate a robust design for AD strategy. The conventional AD design for GFL and SL-GFM with droop control may not provide sufficient damping for droop-I control due to the lack of explicit consideration on nonminimum phase behaviors.
\end{enumerate}





\bibliographystyle{IEEEtranTIE}
\bibliography{IEEEabrv,BIB_manuscript}\ 

\begin{thebibliography}{10}
\providecommand{\url}[1]{#1}
\csname url@samestyle\endcsname
\providecommand{\newblock}{\relax}
\providecommand{\bibinfo}[2]{#2}
\providecommand{\BIBentrySTDinterwordspacing}{\spaceskip=0pt\relax}
\providecommand{\BIBentryALTinterwordstretchfactor}{4}
\providecommand{\BIBentryALTinterwordspacing}{\spaceskip=\fontdimen2\font plus
\BIBentryALTinterwordstretchfactor\fontdimen3\font minus
  \fontdimen4\font\relax}
\providecommand{\BIBforeignlanguage}[2]{{%
\expandafter\ifx\csname l@#1\endcsname\relax
\typeout{** WARNING: IEEEtran.bst: No hyphenation pattern has been}%
\typeout{** loaded for the language `#1'. Using the pattern for}%
\typeout{** the default language instead.}%
\else
\language=\csname l@#1\endcsname
\fi
#2}}
\providecommand{\BIBdecl}{\relax}
\BIBdecl

\bibitem{Chen2024}
M.~Chen, D.~Zhou, A.~Tayyebi, E.~Prieto-Araujo, F.~D\"{o}rfler, and
  F.~Blaabjerg, ``On power control of grid-forming converters: {M}odeling,
  controllability, and full-state feedback design,'' \emph{IEEE Trans. Sustain.
  Energy}, vol.~15, no.~1, pp. 68--80, Jan. 2024.

\bibitem{Pan2020}
D.~Pan, X.~Wang, F.~Liu, and R.~Shi, ``Transient stability of voltage-source
  converters with grid-forming control: A design-oriented study,'' \emph{IEEE
  Trans. Emerg. Sel. Topics Power Electron.}, vol.~8, no.~2, pp. 1019--1033,
  Jun. 2020.

\bibitem{Du2020}
W.~Du, Z.~Chen, K.~P. Schneider, R.~H. Lasseter, S.~Pushpak~Nandanoori, F.~K.
  Tuffner, and S.~Kundu, ``A comparative study of two widely used grid-forming
  droop controls on microgrid small-signal stability,'' \emph{IEEE J. Emerg.
  Sel. Topics Power Electron.}, vol.~8, no.~2, pp. 963--975, Jun. 2020.

\bibitem{Liu2024a}
S.~Liu, H.~Wu, X.~Wang, T.~Bosma, and G.~Sauba, ``Stability analysis and active
  damping design for grid-forming converters in {LC} resonant grids,''
  \emph{IEEE Open J. Ind. Electron. Soc.}, vol.~5, pp. 143--154, 2024.

\bibitem{Wu2023}
H.~Wu and X.~Wang, ``Control of grid-forming {VSCs}: A perspective of adaptive
  fast/slow internal voltage source,'' \emph{IEEE Trans. Power Electron.},
  vol.~38, no.~8, pp. 10\,151--10\,169, Aug. 2023.

\bibitem{Liu2022}
T.~Liu, X.~Wang, F.~Liu, K.~Xin, and Y.~Liu, ``A current limiting method for
  single-loop voltage-magnitude controlled grid-forming converters during
  symmetrical faults,'' \emph{IEEE Trans. Power Electron.}, vol.~37, no.~4, pp.
  4751--4763, Apr. 2022.

\bibitem{Akhavan2023}
A.~Akhavan, J.~C. Vasquez, and J.~M. Guerrero, ``Passivity-based control of
  single-loop grid-forming inverters,'' \emph{IEEE J. Emerg. Sel. Topics Ind.
  Electron.}, vol.~4, no.~2, pp. 571--579, Apr. 2023.

\bibitem{Li2025}
W.~Li, W.~Si, M.~Lu, and J.~Fang, ``Stability analysis and harmonic filtering
  enhancement of single-voltage-loop {PI}- controlled grid-forming
  converters,'' \emph{IEEE Trans. Power Electron.}, vol.~40, no.~4, pp.
  5939--5948, Apr. 2025.

\bibitem{Zhao2022}
F.~Zhao, X.~Wang, and T.~Zhu, ``Power dynamic decoupling control of
  grid-forming converter in stiff grid,'' \emph{IEEE Trans. Power Electron.},
  vol.~37, no.~8, pp. 9073--9088, Aug. 2022.

\bibitem{Sun2023}
P.~Sun, H.~Xu, J.~Yao, Y.~Chi, S.~Huang, and J.~Cao, ``Dynamic interaction
  analysis and damping control strategy of hybrid system with grid-forming and
  grid-following control modes,'' \emph{IEEE Trans. Energy Convers.}, vol.~38,
  no.~3, pp. 1639--1649, Sep. 2023.

\bibitem{Ojo2020}
Y.~Ojo, J.~Watson, and I.~Lestas, ``A review of reduced-order models for
  microgrids: {Simplifications} vs accuracy,'' \emph{arXiv preprint
  arXiv:2003.04923}, 2020.

\bibitem{Liu2025}
R.~Liu, C.~Xue, H.~Zhang, and Y.~R. Li, ``Systematic design of active power
  control parameters for multi-{VSG} systems based on active power
  separation,'' \emph{IEEE Trans. Smart Grid}, vol.~16, no.~4, pp. 3099--3112,
  Jul. 2025.

\bibitem{Rosso2021}
R.~Rosso, X.~Wang, M.~Liserre, X.~Lu, and S.~Engelken, ``Grid-forming
  converters: Control approaches, grid-synchronization, and future trends-{A}
  review,'' \emph{IEEE Open J. Ind. Appl.}, vol.~2, pp. 93--109, May. 2021.

\bibitem{Chen2022}
M.~Chen, D.~Zhou, A.~Tayyebi, E.~Prieto-Araujo, F.~D\"{o}rfler, and
  F.~Blaabjerg, ``Generalized multivariable grid-forming control design for
  power converters,'' \emph{IEEE Trans. Smart Grid}, vol.~13, no.~4, pp.
  2873--2885, Jul. 2022.

\bibitem{Xiong2023}
X.~Xiong, Y.~Zhou, B.~Luo, P.~Cheng, and F.~Blaabjerg, ``Analysis and
  suppression strategy of synchronous frequency resonance for grid-connected
  converters with power-synchronous control method,'' \emph{IEEE Trans. Power
  Electron.}, vol.~38, no.~6, pp. 6945--6955, Jun. 2023.

\bibitem{Parker2014}
S.~G. Parker, B.~P. McGrath, and D.~G. Holmes, ``Regions of active damping
  control for {LCL} filters,'' \emph{IEEE Trans. Ind. Appl.}, vol.~50, no.~1,
  pp. 424--432, Jan./Feb. 2014.

\bibitem{Wang2016}
X.~Wang, F.~Blaabjerg, and P.~C. Loh, ``Grid-current-feedback active damping
  for {LCL} resonance in grid-connected voltage-source converters,'' \emph{IEEE
  Trans. Power Electron.}, vol.~31, no.~1, pp. 213--223, Jan. 2016.

\bibitem{Liu2020}
T.~Liu, J.~Liu, Z.~Liu, and Z.~Liu, ``A study of virtual resistor-based active
  damping alternatives for {LCL} resonance in grid-connected voltage source
  inverters,'' \emph{IEEE Trans. Power Electron.}, vol.~35, no.~1, pp.
  247--262, Jan. 2020.

\bibitem{Wang2022}
X.~Wang, Y.~He, D.~Pan, H.~Zhang, Y.~Ma, and X.~Ruan, ``Passivity enhancement
  for {LCL}-filtered inverter with grid current control and capacitor current
  active damping,'' \emph{IEEE Trans. Power Electron.}, vol.~37, no.~4, pp.
  3801--3812, Apr. 2022.

\bibitem{Franklin2019}
G.~F. Franklin, J.~D. Powell, and A.~Emami-Naeini, \emph{Feedback Control of
  Dynamic Systems}.\hskip 1em plus 0.5em minus 0.4em\relax Pearson, 2019.

\bibitem{Oraa2024}
I.~Oraa, J.~Samanes, J.~Lopez, and E.~Gubia, ``Single-loop droop control
  strategy for a grid-connected {DFIG} wind turbine,'' \emph{IEEE Trans. Ind.
  Electron.}, vol.~71, no.~8, pp. 8819--8830, Aug. 2024.

\bibitem{Sadeque2024}
F.~Sadeque, M.~Gursoy, and B.~Mirafzal, ``Grid-forming inverters in a
  microgrid: Maintaining power during an outage and restoring connection to the
  utility grid without communication,'' \emph{IEEE Trans. Ind. Electron.},
  vol.~71, no.~10, pp. 11\,796--11\,805, Oct. 2024.

\bibitem{Abdelghany2025}
M.~B. Abdelghany, M.~Al~Talaq, S.~Kanukollu, A.~Al-Durra, F.~Gao, and M.~S.~E.
  Moursi, ``Enhanced grid-forming operation of virtual synchronous generator
  units,'' \emph{IEEE Trans. Power Del.}, vol.~40, no.~6, pp. 3588--3603, Dec.
  2025.

\bibitem{Xia2025}
X.~Xia, X.~Zhao, and J.~Liang, ``A novel control strategy to enhance the
  transient performance of grid-forming converters,'' \emph{IEEE Trans. Ind.
  Electron.}, vol.~73, no.~1, pp. 779--790, Jan. 2025.

\bibitem{Chen2021}
M.~Chen, D.~Zhou, and F.~Blaabjerg, ``Voltage control impact on performance of
  virtual synchronous generator,'' in \emph{2021 IEEE 12th Energy Convers.
  Congr. Expo. - Asia (ECCE-Asia)}, Singapore, Singapore, 2021, pp. 1981--1986.

\bibitem{Jayalath2017}
S.~Jayalath and M.~Hanif, ``Generalized {LCL}-filter design algorithm for
  grid-connected voltage-source inverter,'' \emph{IEEE Trans. Ind. Electron.},
  vol.~64, no.~3, pp. 1905--1915, Mar. 2017.

\bibitem{Pooja2024}
Pooja, G.~Teotia, and A.~V.~R. Teja, ``A novel active damping method in grid
  connected inverter using multiple feedback,'' in \emph{IECON 2024 - 50th
  Annu. Conf. IEEE Ind. Electron. Soc. (IECON)}, Chicago, IL, USA, 2024, pp.
  1--6.

\bibitem{Dannehl2010}
J.~Dannehl, F.~W. Fuchs, S.~Hansen, and P.~B. Th{\o}gersen, ``Investigation of
  active damping approaches for {PI}-based current control of grid-connected
  pulse width modulation converters with {LCL} filters,'' \emph{IEEE Trans.
  Ind. Appl.}, vol.~46, no.~4, pp. 1509--1517, Jul./Aug. 2010.

\end{thebibliography}

\end{document}